\documentclass{article}
\newcommand{\arcsec}{\mbox{$^{\prime\prime}$}}
\newcommand{\arcmin}{\mbox{$^{\prime}$}}

\begin{document}
\baselineskip 25pt

\begin{center}

{\bf \Large THE DISTRIBUTION OF REDSHIFTS IN \\
NEW SAMPLES OF QUASI-STELLAR OBJECTS}

\end{center}
\bigskip

\begin{center}
{\large $^\dagger$G.\ Burbidge \& $^\ast$W.M.\ Napier }
\end{center}
\medskip

\begin{center}
$^\dagger${\it Center for Astrophysics and Space Sciences and
Department of Physics, University of California, Mail Code 0424,
San Diego, La Jolla, CA 92093-0424 \\
$^\ast$Armagh Observatory, College Hill, Armagh, BT61 9DG, U.K.}
\end{center}

\vspace{.5in}

\date{Last update 1999 September 6; in original form 1999 September 6}
\newpage
\begin{center}

{\bf \Large ABSTRACT}
\end{center}

\bigskip

Two new samples of QSOs have been constructed from recent surveys to
test the hypothesis that the redshift distribution of bright QSOs is
periodic in $\log(1+z)$.  The first of these comprises 57 different
redshifts among all known close pairs or multiple QSOs, with image
separations $\leq 10\arcsec$, and the second consists of 39 QSOs
selected through their X-ray emission and their proximity to bright
comparatively nearby active galaxies.  The redshift distributions of
the samples are found to exhibit distinct peaks with a periodic
separation of $\sim 0.089 $ in $\log(1+z)$ identical to that claimed in
earlier samples but now extended out to higher redshift peaks $z =
2.63, 3.45$ and 4.47, predicted by the formula but never seen before.
The periodicity is also seen in a third sample, the 78 QSOs of the 3C
and 3CR catalogues. It is present in these three datasets at an
overall significance level $10^{-5}$ - $10^{-6}$, and appears not to be
explicable by spectroscopic or similar selection effects. Possible
interpretations are briefly discussed.


\pagebreak

\section{INTRODUCTION}

Early in the studies of QSOs, a sharp peak in the redshifts at $z = 1.955$
was reported (Burbidge \& Burbidge 1967). Soon after this it was
claimed that if we restrict ourselves to low redshift QSOs and
related objects with similar optical spectra, now called AGN,
the redshifts show a quantized appearance at values of
$z_n = n \times 0.061$, at least up to $n \simeq 10$ (Burbidge, 1968).
Initially, with only 70 objects known, a strong peak was seen at
$z$ = 0.061, and this has persisted with more than 700 objects measured
with $z \leq 0.2$ (Burbidge \& Hewitt 1990).

As the number of redshifts of QSOs grew, additional peaks in the
redshift distribution became apparent. Cowan (1969), Karlsson (1971,
1977) and Barnothy \& Barnothy (1976) claimed that the peaks are
periodic, and Karlsson (1977) found that \mbox{$\Delta$ log$(1 + z) =
0.089$}, with peaks lying at $z$ = 0.061, 0.30, 0.60, 0.96, 1.41, and
1.96.  This analysis was based on about 600 QSOs, most of which were
comparatively bright radio emitting objects (Burbidge, 1978).  The
result was supported by statistical investigations by Fang et
al. (1982) and by Depaquit, Pecker \& Vigier (1985).

The samples showing the periodicity are restricted to quasi-stellar
objects as classically defined, either (a) star-like objects with
large redshifts determined from broad emission lines superimposed on
the blue continua, or (b) objects which although very compact are not
completely stellar in appearance but which have spectra very similar
to those of QSOs. In recent times this latter class have been called
active galactic nuclei or AGN.  The effect has never been found in
samples of galaxies which have normal spectra arising from stars and
interstellar gas, or in samples which have substantial admixtures of
normal galaxies.  Several authors have tested the QSO periodicity
hypothesis using inappropriate samples dominated by types of object
for which the phenomenon has never been claimed. For example Green and
Richstone (1976) failed to find any periodicity in a sample of
emission line galaxies, while Scott (1991) used first all of the
objects identified in the 3CR radio catalogue of which only a small
fraction are QSOs, and then a catalogue of low redshift galaxies
chosen from their IRAS properties (Rowan-Robinson et al. 1990).

Early in the discussion of the peaks and periodicities it was
suggested that instrumental and spectroscopic selection effects
could give rise to spurious peaks at certain redshifts and thus
it has been suggested that the periodicity is simply a reflection
of these effects (e.g. Wills \& Ricklefs 1976, Box \&
Roeder 1984, Scott 1991, Basu 1999). For example, objective prism
surveys yield a large number of QSOs with redshifts between 2 and 2.4,
half-way between the observed peaks.  Depaquit et al. (1985), in an
extensive study of sampling effects, found that there is a strong UV
selection effect in the sampling of optical quasars, leading to the
(U-B) excess of typical QSO spectra peaking at around $z$=0.28 and
1.96, close to two of the postulated peaks, with a weaker maximum at
1.32 (cf Basu 1999). Thus in principle even a radio or x-ray
selected sample in which many sources are identified, color
selection could creep in through a systematic failure to identify
QSOs in the gaps between the peaks.  Where color selection shows
modulation on a scale comparable to the periodicity being claimed
due to filter widths being comparable to emission line spacings,
the reality of the periodicity might well be in doubt.  However,
it was shown long ago that if we take into account all of the
emission lines that are used over the wide spectral range involved
in the redshift determinations it is clear that the observed peaks
are too sharp and too numerous to be explained in this way
(Burbidge 1978).

To improve on the earlier studies we start by considering the well
known statistical association between low redshift galaxies and QSOs
which has been found for many samples.  The first well established
association of this kind was that found between positions of the 3CR
QSOs and the bright galaxies in the Shapley-Ames Catalogue (Burbidge
et al. 1971). Later Burbidge et al. (1990) using a larger sample found
the angular scale of association to be about $\theta \sim 10\arcmin$
corresponding to a separation $\sim$200 kpc ($H_0$ = 60~$\rm
km\,s^{-1}Mpc^{-1}$). Zhu and Chu (1995) found evidence for galaxy -
QSO associations for the Virgo cluster galaxies $5\arcmin \leq
\theta\ \leq 40\arcmin$ consistent with the same linear scale. In very
recent studies (cf Norman \& Williams 2000) similar effects are found.
Having established a statistical association between high redshift
QSOs and low redshift galaxies this provides a new sample of QSOs with
which to test the periodicity effect.  This was tested in 1990
(Karlsson 1990, Arp et al. 1990), and the same periodic effect was
found.  The scale of the associated QSOs was again $\simeq 40
\arcmin.$ Although, in this case the angular separation and the
physical circumstances under which the periodicity is said to hold are
not precisely formulated, the periodicity itself is well-defined and
so, for a prescribed dataset, its presence or otherwise can be
rigorously tested.  Thus in the present paper, we test first the
hypothesis that QSOs close to low-redshift galaxies show a redshift
periodicity $\Delta\log (1+ z)= 0.089$, against the null one that
there is no such periodicity. We then test whether the same
periodicity is present in a second datset chosen on a basis of
morphological characteristics, and finally take a look at the
QSOs in the well-known 3C and 3CR catalogues.

\section{ANALYSIS OF THE KARLSSON DATASET}

In previous studies, little attention has been paid to the standard
deviation of the supposed periodicity, or have mentioned its
phase. However these quantities need to be known if new datasets are
to be used in testing the periodicity claim. Further, the null
distribution is usually implicitly taken to be uniform, although this
is clearly incorrect in general (e.g. Fig.~2).

Here we start by using a bootstrap procedure and apply it to the
Karlsson (1990) data. This yields a narrow confidence region, in
period and phase, within which the periodic solution is plausibly
expected to lie.  The redshift distributions of two new datasets,
described below, are then tested for the Karlsson periodicity, their
significance of fit being given by the fraction of simulated (random)
datasets which yielded a best-fit period and phase in the acceptable
range.  These random datasets are constructed in one or more of three
ways: \\ (i) Data were extracted at random from a distribution of
field QSOs given by the Hewitt \& Burbidge (1993) catalogue of over
7300 QSOs. This catalogue, in common with all other QSO catalogues, is
subject to selection effects. However provided the datasets used for
testing, and those drawn from the catalogue, are equally biased by
these inhomogeneities, the cause of any significant difference between
them must lie elsewhere. \\ (ii) Synthetic data were extracted from
continua constructed so as to empirically reproduce the broadband
behaviour of the data. This procedure empirically allows for the
broadband sampling efficiency involved in the data selection but has
the limitation that narrow-band structure, which might bias the
result, is filtered out. \\ (iii) The real data in the samples were
randomized using a kernel density estimate -- essentially Monte Carlo
sampling with replacement.  This method has the advantage that it
allows one to test whether the `periodicity' is too regular (i.e. the
signal is too strong) to be consistent with shot noise, whatever the
selection function.

The statistic used to investigate the periodicity claim was the power
$I$ of power spectrum analysis (PSA), which has well-known limitations
due to bias, nonstationarity and slowness of convergence to normality
(Newman et al. 1992, Thompson 1990). However since the approach used
is to compare real datasets with an ensemble of synthetic ones, which
suffer these drawbacks in equal measure, there is likely to be little
effect on significance estimates. Conventions vary, but
in the present study the power $I$ is defined as
\[ I(\nu) = 2R^2/N \]
where
\[ R^2=S^2+C^2 \]
with
\[S= \Sigma_{i=1}^N \sin (2\pi x_i/P) \]
\[C= \Sigma_{i=1}^N \cos (2\pi x_i/P) \]
with $\nu=1/P$. For each trial period $P$, the data are in effect
wrapped around a drum of circumference $P$, unit radial vectors ${\bf
e_i}$ are assigned to each datum, and the vectors are added. The
resultant vector
\[{\bf R}= \Sigma_i {\bf e_i}\]
has a direction relative to the x-axis of
\[  \phi = \left\{ \begin{array}{ll}
                     \tan^{-1} (S/C)      & C>0 \\
                     \tan^{-1} (S/C)+ \pi & C<0
                   \end{array}
           \right.
\]
The phase $\phi$ so defined is the distance of the first peak from the
origin along the $\log(1+z)$ axis, and lies in the range 0$\leq\phi
<P$. For random, uniformly distributed redshifts, ${|R|}$ represents a
random walk in the Argand plane, with $\bar{I}$=2, and probability
distribution $\exp(-I/2)$.

Fig.~1a shows a histogram of the 116 QSO redshift data examined by
Karlsson (1990), along with its unwindowed power spectrum (Fig.~1b).
This has a peak in $\Delta \log (1+z)$ corresponding to a periodicity
$(P,\phi)$=(0.089, 0.028) of power $I\sim$15.6.  It is well known that
an unwindowed power spectrum may sometimes throw up high spurious
peaks, and a peak of similar strength appears at $P\sim 0.070$: it
might be thought that the latter should be given similar weight to
that at $\sim 0.089$.  However only the $\sim 0.089$ periodicity had
been previously claimed as real (Karlsson 1971, 1977), and so has to
be given the added weight appropriate to an {\it a priori} claim.

The dispersions in period $P$ and phase $\phi$ obtained from 1000
bootstrap samplings of the data are shown in Fig.~3.  Strong concentrations
of solutions around the two peaks are obvious, as is the presence of
harmonics. It is evident that any new dataset must show
a periodicity in a very narrow range of $(P,\phi)$ before it can be
said to satisfy the hypothesis under test. It was again assumed that
the `real' periodicity, if such exists, is that given by the
well-defined set of solutions in the neighbourhood of the Karlsson
(1990) one, rather than say one of these harmonics.  Thus the
rectangle $ABCD$ defined by $(P,\phi)$= $A$(0.093,0.000),
$B$(0.100,0.000), $C$(0.084,0.060), $D$(0.083,0.036) encompasses 99.5\%
of these neighbourhood solutions and was taken to define the
confidence region within which a best-fit periodicity must lie before
it can be said to be consistent with the hypothesis under test. To a
first approximation, calculated significance levels vary pro rata with
the area of the adopted confidence region.

\section{THE NEW DATASETS}

The first of the two new datasets examined comprises a list of 57
redshifts of all of the known very close pairs or multiple QSOs with
separations between the images $\leq$10~\arcsec (Table~1).  The
majority of objects in this list have been taken from the Castles
survey (CfA-Arizona Space Telescope gravitational lens survey
(Kochanek et al. 2000) and from Table~5 in V\'{e}ron \& V\'{e}ron
(2000). Others have been taken from the recent literature (Lidman et
al. 1999; Myers et al. 1999).  We also include four pairs in which the
two QSOs have very different redshifts (Burbidge et al. 1996, Burbidge
et al. 1997; Surdej et al. 1994; Wampler et al. 1978) which are not in
the Castles list or in Table 5 of V\'{e}ron \& V\'{e}ron.  The QSOs in
Table~1 have usually been put into different physical categories
including: lensed systems, physical pairs, and accidental
configurations. Altogether this table provides a new and unique
sample, since the redshift range is very large (from 0.4 to 4.5) and
many of the QSOs are bright. Practically none of them were known in
the 1970's, so they comprise an independent sample with which to test
the periodicity hypothesis. QSOs with higher redshifts predicted from
the Karlsson series, but not detected in the earlier samples, namely
2.63, 3.45 and 4.47, are present. The distribution of this sample in
\mbox{$\log(1+z)$} is shown in Fig.~4.

In most or all large QSO catalogues, the underlying `complete'
distribution of redshifts (say in a flux-limited sense) is
irredeemably lost due to entanglement with various sampling biases.
For this reason, V\'{e}ron \& V\'{e}ron (2000), for example,
explicitly warn against using their catalogue for statistical
analysis. However, because the periodicity is said to hold only for
QSOs close to galaxies, a differential approach can be applied to this
dataset. Any significant difference between their redshift
distribution and that of field QSOs can then only be due either to
some differential selection effect, or to some physical cause.

The second set is made up of X-ray emitting QSOs lying close to
nearby, active galaxies. It is well known that there are many compact
X-ray sources detected by ROSAT which lie close to active galaxies.
Radecke (1997) showed that there is significant clustering of these
sources about active galaxies, and Arp (1997) identified many of them
as QSO candidates.  All of the objects which have so far been observed
spectroscopically turn out to be QSOs. These 39 QSOs are listed in
Table~2. The results are taken from Burbidge (1995, 1997, 1999a,
1999b), Burbidge et al (1999), Arp (1996, 1997) and Chu et al. (1998).
This sample is `complete' in the sense that all the compact X-ray
sources close to nearby, active galaxies were identified as QSOs in
the ROSAT survey to the prescribed flux limit. There are therefore no
`missing' QSOs in this sample which could allow gaps between peaks to
be filled. This redshift distribution is shown in Fig.~5.

\subsection {The close pairs of QSOs}

Power spectrum analysis of the redshifts in Table~1 yields a clear
peak of signal strength $I\sim$10.6 at $(P,\phi)$=(0.093, 0.007).

To a first approximation, the QSOs in this list are simply the lensed
fraction of the brightest QSOs in (say) the Hewitt \& Burbidge (1993)
catalogue, modulated by some smooth function to take account of the
lensing probability (which is a function of magnitude). There is
therefore no reason to expect high-frequency modulation in the dataset
other than that which already exists in the QSO catalogues.  We first
ask whether this signal might arise by chance, from that of QSOs in
the general field.  To assess this, data were randomly extracted, with
replacement, from the redshift distribution of the $\sim$7300 QSOs in
the Hewitt \& Burbidge (1993) catalogue (Fig.~2) in sets of 57 and
analyzed as above, the peak values of $(P,\phi)$ in the power spectrum
of each dataset being recorded.  The distribution of 5,000 such peaks
is shown in Fig.~6. It has a banded structure, which may be ascribed
to edge effects (Lutz 1985), but the peaks clearly extend over the
whole range of $(P,\phi)$ investigated. For a run of
5,000 such trials, 87 peaks occurred within the prescribed confidence
region, and of those, 13 had power $I\geq$10.6. Thus the probability
that a set of 57 QSOs, randomly extracted from the Hewitt-Burbidge
catalogue, would yield the observed periodicity by chance is $\sim
2.6\times 10^{-3}$, and the dataset under examination confirms the
Karlsson periodic solution at a confidence level 99.7\%. The
lensing probability varies smoothly with magnitude and
so is unlikely to have much effect on the calculated significance.

A second test was carried out in which 57 redshifts were extracted
from the dataset at random, with replacement, and a random gaussian
displacement applied to each, with dispersion $\sigma$=7 km\,s$^{-1}$
sufficient to wash out the periodicity but not enough to mask the
overall redshift distribution. PSA was applied to each dataset so
constructed and the highest peak occurring anywhere in the range
0.040$\leq P\leq$0.200 was recorded. A run of 5,000 such trials
yielded 104 `hits' within the ABCD quadrilateral, of which 17 had
power $I\geq$10.6. Thus these Monte Carlo trials once again support
the Karlsson periodic solution at a confidence level $C\sim$99.7\%.

\subsection {The X-ray QSOs}

A PSA of the 39 X-ray redshifts yields a formal solution $(P,\phi)$=(0.083,
0.047), just within the Karlsson solution but with strength only
$I\sim$7.2. It is unlikely that the X-ray QSOs represent a sample
drawn from a distribution like that of Fig.~2 and so, to assess the
significance of this best-fit periodicity, a null distribution was
derived by fitting a cubic spline on to the data
(Fig.~8).  Synthetic data were then extracted randomly in
sets of 39 from this continuum, and PSA applied to them, the highest
peaks anywhere in the range 0.040$\leq P\leq$0.200 being
recorded. Because of
the broadband structure of this null continuum, the periodicity is
here being tested against a `shot noise' null hypothesis.
This procedure yielded 90 `hits', 36 of them
with $I\geq$7.2, whence $C\sim$99.2\%. A similar procedure using
datasets generated by the Monte Carlo procedure yielded
132 hits, 53 of them with $I\geq$7.2, whence $C\sim$98.9\%.

\subsection {The combined datasets}

The two datasets are almost complementary in that most of the
redshifts in Table~1 have $z\geq 1.41$ while most of the `X-ray QSO'
redshifts in Table 2 have $z\leq 1.41$. The combined samples amount to
a set of 96 redshifts, similar in size to the 116 Karlsson (1990)
dataset being used as a template. Their distribution in
\mbox{$\log(1+z)$} is shown `raw' in Fig.~9a, and `smoothed' in
Fig.~9b. This latter was obtained from a rectangular window, which is
an asymptotically unbiased kernel estimator (de Jager et al. 1986), of
width 0.003 and step length 0.001. A periodicity is clearly visible,
the 96 combined data yielding a peak signal at
$(P,\phi)$=(0.086,0.039) of strength $I\sim$15.6 (Fig.~10). It would,
of course, be surprising if this pooling of data introduced
periodicity where none exists in the individual datasets. This
best-fit solution is remarkably close to
the one under test, namely $(P,\phi)$=(0.089, 0.028).

To test whether there is a significant difference between these
solutions, the data from the Karlsson and two new datasets were first
combined.  A PSA of the 212 redshifts yielded
$(P,\phi)$=(0.088,0.034), the signal having strength $I$=29.5.  A
Monte Carlo $t$-test was then carried out in which the 96 redshifts of
Tables~1 and 2 were first pooled with the 116 of the Karlsson dataset,
and then randomly reallocated into pairs of (116,96). For each such
pair, the mean residual $\bar{x_i}$ departure from the solution
$(P,\phi)$=(0.088,0.034) was calculated, and the difference
$\bar{x_1}-\bar{x_2}$ recorded. Three thousand such trials (Fig.~7)
yielded a basis for comparison with the observed
$\bar{x_1}-\bar{x_2}$=0.015. It can be seen that there is no
significant difference between the best-fit periodicity derived from
the new datasets and the one under test.

The significance of this periodic solution was estimated in a number
of ways. Synthetic datasets were created by the Monte Carlo
procedure as before: randomly choosing data
(with replenishment) and then adding, to each datum so chosen, a
random gaussian element.  A dispersion $\sigma$=0.090 was first
adopted, large enough to wash out the periodicity under test but not
so large as to wash out the overall distribution of the dataset.  The
overall significance level was given by the fraction of solutions
lying within the confidence region, having at least the observed
signal strength. A set of 5000 trials yielded 114 `hits' within the
prescribed confidence region, of which three had $I\ge$15.5
(Fig.~11). Similar runs with $\sigma$=0.060 and 0.030 yielded similar
answers. These significance levels were checked by creating a null
continuum with a spline fit as before, extracting data in sets of 96 and
recording the best-fit $(P,\phi)$, and repeating the operation 5,000
times. The results were essentially identical. The trials imply that
the periodicity has significance $\sim10^{-4}$: this
significance level for the datasets in combination is consistent with
those obtained from them individually.

However, when the dispersion was dropped to $\sigma$=0.01 in the Monte
Carlo runs, a little over a tenth of the periodicity, the behaviour
changed dramatically: a run of 5000 trials yielded 2829 `hits' of
which 1390 had $I\ge$15.5. That is, the dispersion must virtually
disappear before the high frequency signal is seen. This result
implies that the signal is not due to (say) color selection effects
which simply modulate on `characteristic scales' comparable with the
separations between peaks. The modulation, whether due to selection
effects or `new physics', must be periodic, with a peak-to-peak dispersion
not much greater than $\sim 0.1P$.

In interpreting these significance levels, account should be taken of
the fact that the hypothesis has not been formulated with precision in
the literature (what proximity cutoff is appropriate?). In general the
freedom to make choices will reduce formal significance levels. This
effect is difficult to quantify, but in the present study no such
freedoms were exercised, and it is likely that the overall reduction
in significance is small.

\section{RADIO-SELECTED QSOS}

It was pointed out earlier that nearly all of the QSOs in which the peaks and periodicity were originally found
were radio sources.  However those samples were not subjected to statistical analysis of the kind used here.
Thus we have considered it worthwhile to take such a sample and carry out the analysis.
The sample we have chosen is made up of the QSOs originally identified from the 3C radio
catalogue and after it was revised, from the 3CR catalogue.  This is a complete catalogue of sources covering
about 2/3 of the sky listing the most powerful radio sources measured
at 178 MHz. The optical objects identified with the radio sources are
either normal galaxies, broad emission line radio galaxies or QSOs.
We are only concerned here with the QSOs.  There are 78 of these, 24
are 3C sources, and 54 are 3CR sources. The data are shown in
Table~3. The redshifts and magnitudes have been taken from the QSO
catalogue of Hewitt \& Burbidge (1993).  Only two of all of the
objects morphologically classified as QSOs have been omitted, 3C82
whose redshift is uncertain, and 3CR371 which is a BL Lac object.

The histogram of the redshift distribution of the 78 is shown in
Fig~12, along with its power spectrum. The signal $I\sim$5.7 is very
weak, but the peak at $(P,\phi)$=(0.091,0.028) lies centrally within
the confidence region of Fig.~3.  Trials on synthetic datasets
(obtained by Monte Carlo sampling with replenishment, as described
above) reveal that a periodic signal of this strength and $(P,\phi)$
is present with probability $\sim$98.5\%. Thus
the evidence is that the 3C and 3CR data also reveal the phenomenon.

The three datasets combined yield $(P,\phi)$=(0.087,0.034) with
signal strength $I\sim$18.6. Five thousand Monte Carlo simulations yielded
no synthetic datasets in the confidence region with this signal strength.

\section{INTERPRETATION}

These results, from quite different datasets, indicate
that the redshift distribution of QSOs in the
datasets examined are not consistent with random sampling from the
general QSO field. A periodicity is seen (Fig.~9): it is identical,
within the uncertainties, to that which has now been reported several
times in the literature for QSOs close to galaxies. It is remarkably
regular and is found to extend for at least two cycles beyond its
previously known range. It has been found to be present in: \\
(i) QSOs close to companion galaxies; \\
(ii) QSOs which, on investigation, turn out to be binary or
multiple; \\
(iii) X-ray sources close to bright active galaxies which,
on investigation, turn out to be QSOs; and \\
(iv) the 3C and 3CR QSOs, which effectively comprise a complete sample.

It is difficult to see how an artefact (say generated through color or
emission-line selection) could achieve this.  The latter two samples
in particular have been constructed in such a way that color effects
are effectively excluded.  Further, any such artefact would also have
to imitate a periodicity to a high degree of regularity (Fig.~9): a
`characteristic modulation' does not reproduce the behaviour of the
datasets. Thus, we now assume that the periodicity is physically real and
discuss some possible interpretations.

In general, for any extragalactic object, the observed redshift $z_0$
can be written in the form

(1 + $z_0$) = (1 +$z_c$)(1+$z_d$)(1+$z_i$)

where $z_c$, $z_d$, and $z_i$ respectively are redshift components due
to the expansion of the universe, random motions (sometimes called
peculiar velocities or other velocities) and intrinsic properties
(associated with the physics of the objects).  If the redshifts of the
QSOs have small $z_d$ and $z_i$, so that $z_0 \simeq z_c$, then the
observed periodicity must imply some kind of oscillating behavior in
the evolution of the universe. Such oscillations are a generic feature
of many models with scalar-tensor gravity (e.g. Busarello et
al. 1994), and in the case of a vacuum-dominated Universe they have a
$\log (1+z)$ periodicity. Might the periodicity then be an
inflationary remnant? If this were the case all extragalactic objects
including normal galaxies should show the same effects, and there is
no evidence that they do.

We have a situation in which the periodicity is confined to QSOs
close to other QSOs or close to active galaxies.  For the sample
based on the association of X-ray QSOs with bright galaxies, and
the sample of Karlsson, the galaxies are all very closeby and thus
$z_c$ is very small (cf Table 2).

Thus the most probable interpretation for these two samples is that in
each case $z_0 \simeq z_i$, $z_c$ is very small and the value of $z_i$ for each QSO is one
of the values 0.061, 0.30, 0.60, 0.96, 1.41, 1.96, 2.63, 3.45 or 4.47.  Thus for the
QSOs in Table 2 we can calculate the values of $z_d$.  It turns out
that for these QSOs there are both redshifted and blueshifted values
of $ z_d$ and the average value  $|cz_d| \simeq$12500~km\,s$^{-1}$. This may
be a measure of the line-of-sight velocity component associated with
the ejection of the QSO from the galaxy.  A similar argument applies
to a number of the 3CR QSOs shown in Table 3.  It was shown long ago
that several of these QSOs (3C 232, 3CR 268.4, 3CR 275.1, 3CR 309.1,
3CR 345 and 3CR 455) lie very close to nearby bright galaxies, and
statistical studies and some morphological studies strongly suggest
that they are physically associated with those galaxies (Burbidge et
al. 1971; Burbidge 1996).

What about the QSOs in Table 1?  The general view has been that many
of them are very distant as is required for the gravitational lens
interpretation, and that the observed redshifts are totally
cosmological in origin.  However, the fact that they show the same
peaks and periodicity predicted from the lower redshift objects, which
we have shown are comparatively nearby, may cast doubt on this
classical interpretation.

In this connection it is also worth pointing out that: \\(i) The first
gravitational lens candidate ever discovered, 0957+561A\&B, which is
listed in Table 1, has $z_0 = 1.41$, exactly on a peak, and it is an
X-ray source lying only $15^{\prime}$ from the nearby active galaxy NGC 3079.
This object is included in Table 2 so that it is the one system in
both data sets.  The possible physical association of 0957 + 561 with
NGC 3079 has been ignored by those who believe that it is a lensed
object. \\(ii) The redshifts in Table 1 are from objects which have
been chosen from their morphological characteristics alone, i.e.,
their multiplicity.  Thus in that Table there are 18 out of the 57
redshifts or about 30 percent arising from objects that are \underline
{not} gravitational lens candidates.  They are either called binary
systems, or they are very close pairs with very different redshifts, the latter
already suggesting that perhaps the redshifts are not of cosmological origin
(cf Burbidge et al. 1997).  Thus for a significant fraction of the
objects in Table 1 the argument cannot be made that there is a
conflict between the results obtained here and the gravitational lens
hypothesis.  Even among the gravitational lens candidates there are
only 12 in which it is claimed that the lensing galaxy has been
identified and its redshift measured (V\'{e}ron \& V\'{e}ron 2000,
Table 5).

The results obtained in this paper together with the earlier work,
and the statistical evidence for associations between galaxies
with comparatively small redshifts and QSOs with large redshifts
suggest that QSOs with intrinsic components are ejected from
galaxies. If the sample is comparatively nearby so that $z_c$ is
very small the intrinsic redshift $z_i$ will dominate and this
periodic effect will be seen.  For QSOs ejected from galaxies with
non-negligible values of  $z_c$ the periodicity will not be seen
because of the smearing due to the cosmological term.  However for
properly chosen samples of QSOs and galaxies the clustering
tendencies will still be detectable.

We are indebted to Margaret Burbidge and Fred Hoyle for many
helpful discussions.

\newpage
\begin{center}
\bf{REFERENCES}
\end{center}
\begin{description}
\vspace{.5in}

\item[] Arp, H.C., 1996, A\&A, 316, 57

\item[] Arp, H.C., 1997, A\&A, 319, 33

\item[] Arp, H.C., Bi, H., Chu, Y. \& Zhu, X. 1990, A\&A, 239, 33

\item[] Basu, D., 1999, Astron. Nachr., 320, 53

\item[] Barnothy, J. \& Barnothy, M., 1976, Pub. A.S.P., 88,
837

\item[] Box, T.C. \& Roeder, R.C., 1984, A\&A, 134, 234

\item[] Burbidge, E.M., Beaver, E., Cohen, R.D., Junkkarinen,
V. \& Lyons, R., 1996, A. J., 112, 2533

\item[] Burbidge, E.M., 1995, A\&A, 298, L1

\item[] Burbidge, E.M., 1997, ApJ., 484, L99

\item[] Burbidge, E.M., 1999a, ApJ., 511, L11

\item[] Burbidge, E.M., 1999b, Private Communication

\item[] Burbidge, E.M., Arp, H. \& Chu, Y. 2000, Preprint

\item[] Burbidge, E.M., Burbidge, G., Solomon, P., \&
Strittmatter, P., 1971, ApJ., 170, 223

\item[] Burbidge,E. M. \& Burbidge, G., 1967, ApJ., 148, L107

\item[] Burbidge, G., 1996, A\&A, 309, 9

\item[] Burbidge, G., 1968, ApJ. 154, L41

\item[] Burbidge, G., 1978, Phys. Scripta, 17, 237

\item[] Burbidge, G., \& Hewitt, A., 1990, ApJ., 359, L33

\item[] Burbidge, G., Hewitt, A., Narlikar, J. V. \& Das Gupta, P.
1990, ApJS, 74, 675

\item[] Burbidge, G., Hoyle, F., \& Schneider, P., 1997, A\&A,
320, 8

\item[] Busarello, G., Capozziello, S., de Ritis, R., Longo, G.,
Rifatto, A., Rubano, C. \& Scudarello, P., 1994,
A\&A, 283, 717

\item[] Chu, Y., Wei, J., Hu, J., Zhu, X. \& Arp., H.C., 1998,
ApJ., 500, 596

\item[] Cowan, C. L., 1969, Nature, 224, 655

\item[] Depaquit, S., Pecker, J.C., \& Vigier, J-P., 1985, Astron.
Nachr., 306, 7

\item[] Fang, L., Chu, Y., Lin., Y., 1982, A\&A, 106, 287

\item[] Green, R. \& Richstone, D., 1976, ApJ., 208, 639

\item[] Hewitt, A. \& Burbidge, G., 1993, ApJS., 87, 451

\item[] Karlsson, J.K., 1971, A\&A, 13,333

\item[] Karlsson, J.K., 1977, A\&A, 58, 237

\item[] Karlsson, J.K., 1990, A\&A, 239, 50

\item[] Kochanek, C. K., Falco, E., Impey, C. Lehar, B.,
McCleod, R., \& Rix, S., 2000, http:-cfa-www.harvard.edu/castles.

\item[] Lidman, C., Courbin, F., Meylan, G., Broadbent, T.,
Frye, B., \& Welch, J., 1999, ApJ., 514, L57

\item[] Lutz, T.M., 1985, Nature 317, 404.

\item[] Myers, S. T., et al., 1999, A.J., 117, 2565

\item[] Norman \& Williams, A.J., 119, 2060, 2000

\item[] Newman, W.I., Haynes, M. P. \& Terzian, Y., 1992,
``Statistical Challenges in Modern Astronomy", (Ed. E. Feigelson
\& G. Babu, Springer Verlag, New York) p. 137

\item[] Radecke, H. D., 1997, A\&A, 319,18

\item[] Rowan-Robinson, M., et al., 1990, MNRAS, 247, 1

\item[] Scott, D., 1991, A\&A, 242, 1

\item[] Surdej, J., et al., 1994, ``Gravitational Lenses in the
Universe", 31st Liege Colloquium Proc.(Ed. J. Surdej et al.) p.
153

\item[] Thompson, D.J., 1990, Phil. Trans Roy. Soc., London, A
330, 601

\item[] Veron, P. \& Veron, M., 2000, ESO Scientific Report No.
18

\item[] Wampler, E. J., Baldwin, J., Burke, W., Robinson, L. \&
Hazard, C., 1978, Nature, 246, 203

\item[] Wills, D. \& Ricklefs, R. L., 1976, MNRAS, 175, 65P

\item[] Zhu, X-F. \& Chu, Y-Q, 1995, A \& A 297, 300
\end{description}

\newpage
\begin{description}

\item[Fig. 1]
(a) Histogram of the 116 Karlsson QSO redshift data,
plotted in units of 100 $\log (1+z)$. The vertical lines represent the
periodicity peaks claimed by Karlsson. (b) Unwindowed power spectrum
of the data. Power $I$ is defined so that $\bar{I}$=2 for white noise.

\item[Fig. 2]
Distribution of QSO redshifts from the Burbidge \& Hewitt (1993)
catalogue, in units of 100 $\log (1+z)$.

\item[Fig. 3]
Disperson in $(P,\phi)$ from bootstrap sampling of the Karlsson data.

\item[Fig. 4]
Histogram of the redshift distribution of close pairs and multiple
QSOs plotted in units of 100$\log(1+z)$, stepping in intervals of 1.
The vertical dotted lines represent the positions of the peaks under test.

\item[Fig. 5]
The redshift distribution of X ray-selected QSOs
plotted using the same scales as in Fig. 4.

\item[Fig. 6]
The $(P,\phi)$ distribution of peaks obtained by random extraction
from the data plotted in Fig.~4.

\item[Fig. 7] A $t$-test to determine whether the best-fit periodicity
$(P,\phi)$=(0.086,0.039), obtained for the two new datasets, differs
significantly from the periodicity $(P,\phi)$=(0.089,0.028) under
test.  For each trial, the 116 Karlsson and 96 (Table~1+2) redshifts
are randomly reallocated into two sets of 116 and 96 redshifts, and
for each subset the mean of the residuals $\bar{x_i}$ from the
prescribed best-fit periodicity is calculated. The differences
$\bar{x_1}-\bar{x_2}$ are plotted for 3000 trials, as shown. For the
real datasets, $\bar{x_1}-\bar{x_2}$=0.015.

\item [Fig. 8]
Cubic spline fit to the X-ray data plotted in Fig.~5.

\item[Fig. 9]
The combined redshift distribution of the QSOs in Tables 1 and 2.
(a) Raw plot in units of  100 $\log (1+z)$.
(b) Data smoothed with a running window of width 3 units.

\item[Fig. 10]
Power spectrum of the combined dataset. The peak has $I\sim$14.6 at
$\Delta \log (1+z)$=0.088 and phase 0.031.

\item[Fig. 11]
$(P,\phi)$ distribution for 5,000 synthetic datasets simulating the
Fig.~9 distribution.

\item[Fig. 12]
(a) The redshift distribution of
3C and 3CR radio-selected QSOs in Table~3. \\
(b) The corresponding power spectrum.

\end{description}
\pagebreak

\newcommand{\spa}[1]{\Huge\renewcommand{\baselinestretch}{#1}\normalsize}
\spa{1}

\begin{center}
{\bf TABLE 1\\
 Binary and multiple QSOs with  separations $<$10\arcsec}
\end{center}

\begin{tabular}{r|r|r|r|r}
 Object             & $z$       &  $m$  &  Sepn.   & No. of \\
                   &           &       & (\arcsec) & components \\
                   \hline\hline
 MG 0023+171        & 0.95     &             & 4.8       & 2  \\
 0047-2808          & 3.60     &  23.9       & 2.7       &    \\
 UM673=0142-100     & 2.72  & 16.8V       & 2.2       & 2  \\
 PHL1222=0151+048   & 1.91     & 17.63V      & 3.3       &    \\
 CTQ414=0158-4325   & 1.29     &             & 1.2       & 2  \\
 B 0218+357         & 0.96     & 20.0V       & 0.33      & 2  \\
 HE 0230-2130       & 2.16     & 18.2V       & 2.0       & 3  \\
0235+164A          & 0.94     & av. 18      & 2.5       & 1  \\
 0235+164B          & 0.52     & 19          & 2.5       & 1  \\
 QJ 0240-343        & 1.41     & 20.4V       & 6.1       & 2  \\
 PKS MG 0414+0534   & 2.64     & 19.3,21.3R  & 2.2       & 4  \\
 B 0172+472         & 1.34     & 23.0V       & 1.27      & 4  \\
MG 0751+2716       & 3.20     &             & 0.9       &    \\
APM 08279+5255     & 3.87     & 15.2        & 0.4       & 2  \\
 SBS 0909+532       & 1.38 & 17.0,17.24V & 1.1       & 2  \\
 RXJ 0911.4+0551    & 2.80  & 18.34 V     & 0.8(3.1)  & 4  \\
FBQ 0951+2635      & 1.24     & 16.9        & 1.1       & 2  \\
 BRI 0952-0115      & 4.43     & 18.9R       & 0.95      & 2  \\
Q 0957+561         & 1.41     & 16.7        & 6.1       & 3  \\
 LBQS 1009-0252     & 2.74     & 18.2        & 1.55      & 2  \\
 1009-0252C         & 1.62     & 19.3        &           &    \\
 J03.13=Q1017-207   & 2.55     & 17.1        & 0.86      & 2  \\
 (1015-20)          &          &             &           &    \\
 FSC 10214+4724     & 2.29     & 20.5R       & 1.18      & 2  \\
 B 1030+074         & 1.54     & 20.34V      & 1.56      & 2  \\
 HE 1104-1805       & 2.32     & 16.9        & 3.0       & 2  \\
 PG 1115+080        & 1.72     & 15.8        & 2.3       & 4  \\
 UM425=Q 1120+0195  & 1.46     & 16.1V       & 6.5       & 2  \\
 PKS 1145-071       & 1.35     & 18V         & 4.2       & 2  \\
 1148+055A          & 1.89     & 17.9        & 3.9       & 1  \\
 1148+055B          & 1.41     & 20.7        & 0.45      & 1  \\
 B1152+199  &  1.02&16.5&1.6&1\\
 1208+1011          & 3.80     & 18.1        & 8.9       & 2  \\
 HS 1216+5032       & 1.45     & 17.2V,19.1V &           & 2  \\
RRSIV 27Q 1343+2640 & 2.03     & 20.0        & 9.5       & 2  \\
 B1359+154&3.24&22&$< 1.7$\\
 HST 1413+117       & 2.55     & 17          & 1.4 (1.2) & 4  \\
 HST 14176+5226     & 3.40     & 24.3?       & 3.2 (1.4) & 4  \\
 B 1422+231         & 3.62     & 15.6        & 1.3       & 4  \\

\end{tabular}

\newpage

\begin{center}
{\bf TABLE 1 cont.}
\end{center}
\medskip

\begin{tabular}{r|r|r|r|r}
Object             & $z$       &  $m$  &  Sepn.   & No. of \\
                 &           &       & (\arcsec) & components \\
                 \hline\hline
LBQS 1429-008      & 2.08     & 17.7        & 5.1       & 2      \\
 SBS 1520+530       & 1.86     & 18.2        & 1.6       & 2      \\
 1548+114A          & 1.90     & 18.1        & 4.8       & 1      \\
                  1549+114B          & 0.44     & 18.8        & 4.8       & 1      \\
 B 1600+434         & 1.59  & 20R      & 1.38      & 2      \\
B 1608+656         & 1.39     & 20R         & 2.1       & 4      \\
FBQ 1633+3134      & 1.52     &             & 0.66      & 2      \\
 Q 1634+267         & 1.96  & 18.5, 20.0        & 3.8       & 2      \\
 J 1643+3156        & 0.59     & 18.4, 19.2   & 2.3       & 2      \\
 MG 1654+1346       & 1.74     & 20.9R       & 2.1       &       \\
PKS 1830-211       & 2.51     &             &           & 2      \\
 MG 2016+1127       & 3.27     & 22.1E       & 3.4 (3.8) & 3      \\
 2045+265           & 1.28     &             & 1.86      & 4      \\
 Q 2138-431         & 1.64     & 18.85V      & 4.5       & 2      \\
 HE 2149-2745       & 2.03     & 17.3B       & 1.7       & 2      \\
 LBQS 2153-2056     & 1.85 &             & 7.8       & 2      \\
 MGC 2214+3550      & 0.88     & 19.3        & 3.0       & 4      \\
 2237+0305          & 1.69     & B16.8       & 1.8       & 4      \\
 Q 2345+007 A,B     & 2.15     & 19.5, 21.0 & 7.1       & 2      \\

\end{tabular}
\pagebreak
\spa{1}

\begin{center}
{\bf TABLE 2} \\X-ray selected QSOs close to bright, active galaxies
\medskip
\end{center}

\begin{tabular}{r|r|r} \\
Galaxy   & $z_{gal}$ & $z_{Q}$ \\ \hline\hline
NGC 1068 & 0.004  & 0.261 \\
         &        & 0.388 \\
         &        & 0.655 \\ \hline
NGC 2639 & 0.011  & 0.305 \\
         &        & 0.323 \\ \hline
NGC 3079 & 0.0038 & 0.216 \\
         &        & 1.022  \\
         &        & 1.41  \\ \hline
NGC 3516 & 0.0087 & 0.328 \\
         &        & 0.690 \\
         &        & 0.929 \\
         &        & 1.399 \\
         &        & 2.10  \\ \hline
NGC 3628 & 0.003  & 0.983 \\ \hline
NGC 4258 & 0.0015 & 0.398 \\
          &       & 0.653 \\ \hline
NGC 4579 & 0.005  & 0.106 \\
          &        & 0.662 \\
          &        & 0.947 \\ \hline
Mkn 231  & 0.041  & 0.320 \\
         &        & 0.489 \\ \hline
Mkn 273  & 0.038  & 0.376  \\
         &        & 0.600 \\
         &        & 0.941 \\
         &        & 1.163 \\ \hline
NGC 5273 & 0.0035 & 0.336  \\ \hline
NGC 5548 & 0.017  & 0.184 \\
         &        & 0.560 \\
         &        & 0.674 \\
         &        & 0.727 \\
         &        & 0.852 \\ \hline
NGC 5689 & 0.0076 & 1.358 \\
         &        & 1.94  \\
         &        & 2.391 \\ \hline
IC 4553  & 0.018  & 0.459 \\ \hline
NGC 6217 & 0.005  & 0.358 \\
          &        & 0.376 \\
          &        & 0.380 \\
          &        & 1.134 \\

\end{tabular}
\pagebreak

\spa{1}

\begin{center}
{\bf TABLE 3}
\end{center}

\begin{tabular}{l|l|l|l}

Object&Name &$m_v$&$z$\\
\hline\hline
0219+428&3C 66A&15.58&0.444\\
0350-073&3C 94&17.73&0.962\\
0349-146&3C 95&16.22&0.616\\
0409+229&3C 108&18.7&1.215\\
0414-060&3C 110&16.25&0.781\\
0723+679&3C 179&18.0&0.846\\
0736-019&3C 185&17.6&1.033\\
0814+227&3C 197&18&0.98\\
0837-120&3C 206&15.76&0.198\\
0955+326&3C 232&15.78&0.533\\
1015+277&3C 240&17.5&0.469\\
1023+067&3C 243&18.54&1.707\\
1048-090&3C 246&16.79&0.344\\
1132+303&3C 261&18.24&0.614\\
1253-055&3C 279&16.84&0.538\\
1305+069&3C 281&17.02&0.602\\
1441+522&3C 303C&19.97&1.57\\
1502+602&3C 311&18&1.022\\
1634+269&3C 342&17.75&0.988\\
1901+319&3C 395&17.42&0.635\\
2005-044&3C 407&18&0.589\\
2044-027&3C 422&19.5&0.942\\
2223-052&3C 446&17.19&1.404\\
2325+269&3C 463&17.5&0.875\\
\hline\hline
0003-003&3CR 2&19.35&1.037\\
0017+154&3CR 9&18.21&2.018\\
0033+183&3CR 14&20&1.469\\
0127+233&3CR 43&20&1.459\\
0133+207&3CR 47&18.1&0.425\\
0134+329&3CR 48 &16.46&0.367\\
0141+339&3CR 48/54&17.01&1.455\\
0210+860&3CR 61.1&19&0.184\\
0229+341&3CR 68.1&19&1.238\\
0340+048&3CR 93&17.73&0.357\\
0518+165&3CR 138&18.84&0.759\\
0538+498&3CR 147&17.8&0.545\\
0610+260&3CR 154&18&0.580\\
0710+118&3CR 175&16.6&0.768\\
0725+147&3CR 181&17.68&1.387\\

\end{tabular}
\pagebreak
\spa{1}

\begin{center}
{\bf TABLE 3 cont.}
\end{center}

\begin{tabular}{l|l|l|l}
Object&Name&$m_v$&$z$\\
\hline\hline

0740+380&3CR 186&17.6&1.063\\
0758+143&3CR 190&20.32&1.195\\
0802+103&3CR 191&18.19&1.956\\
0809+483&3CR 196&17.79&0.871\\
0833+654&3CR 204&18.21&1.112\\
0835+580&3CR 205&17.62&1.536\\
0838+133&3CR 207&18.15&0.684\\
0850+140&3CR 208&17.30&1.11\\
0855+143&3CR 212&19.06&1.048\\
0903+169&3CR 215&18.27&0.411\\
0906+430&3CR 216&18.48&0.67\\
0927+362&3CR 220.2&19&1.157\\
1040+123&3CR 245&16.45&1.029\\
1100+772&3CR 249.1&15.72&0.311\\
1111+408&3CR 254&17.98&0.734\\
1137+660&3CR 263&16.32&0.652\\
1206+439&3CR 268.4&18.42&1.396\\
1218+339&3CR 270.1&18.61&1.516\\
1226+023&3CR 273&13.02&0.158\\
1241+166&3CR 275.1&19&0.557\\
1250+568&3CR 277.1&17.93&0.321\\
1258+404&3CR 280.1&19.44&1.667\\
1328+307&3CR 286&17.25&0.849\\
1328+254&3CR 287&17.67&1.055\\
1340+606&3CR 288.1&18.12&0.961\\
1416+067&3CR 298&16.79&1.439\\
1458+718&3CR 309.1&16.78&0.905\\
1545+210&3CR 323.1 &16.69&0.264\\
1618+177&3CR 334&16.77&0.555\\
1622+238&3CR 336&17.47&0.927\\
1634+628&3CR 343&20.6&0.988\\
1641+399&3CR 345&15.96&0.595\\
1704+608&3CR 351&16.01&0.371\\
1828+487&3CR 380&16.81&0.692\\
2037+511&3CR 418&20&1.686\\
2120+168&3CR 432&17.96&1.805\\
2249+185&3CR 454&18.47&1.761\\
2251+158&3CR 454.3&16.1&0.859\\
2252+129&3CR 455&19.7&0.543\\

\end{tabular}

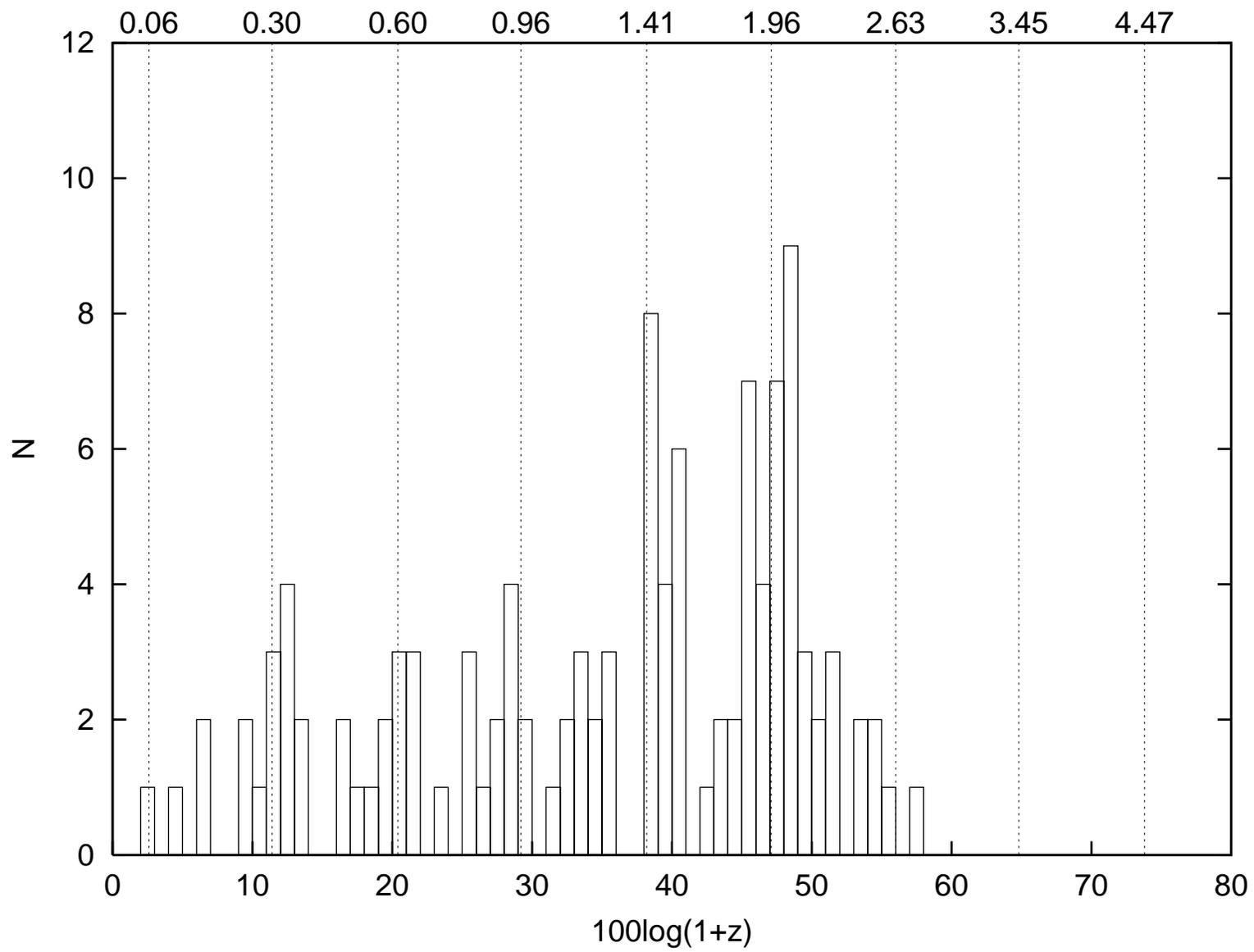
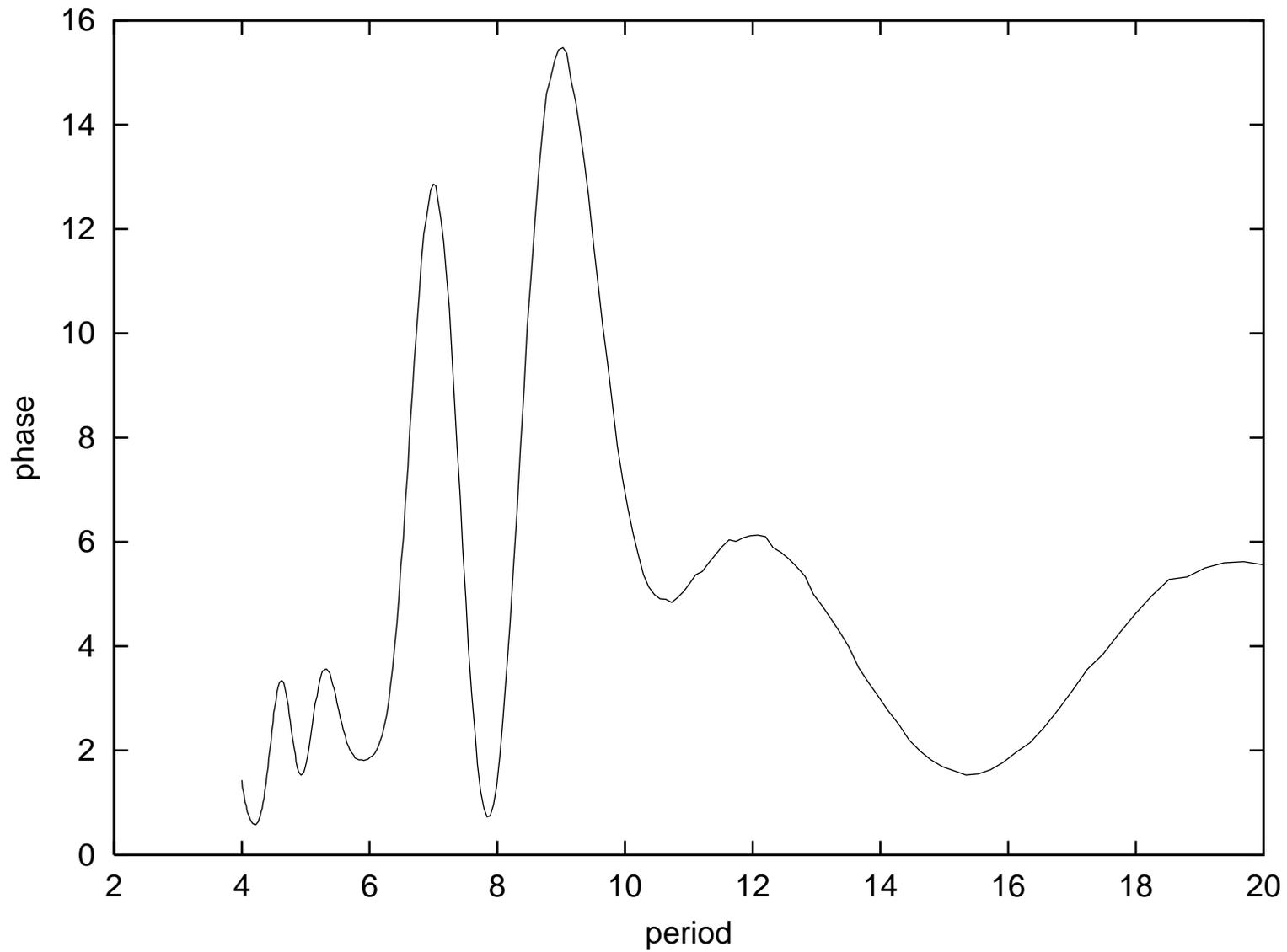
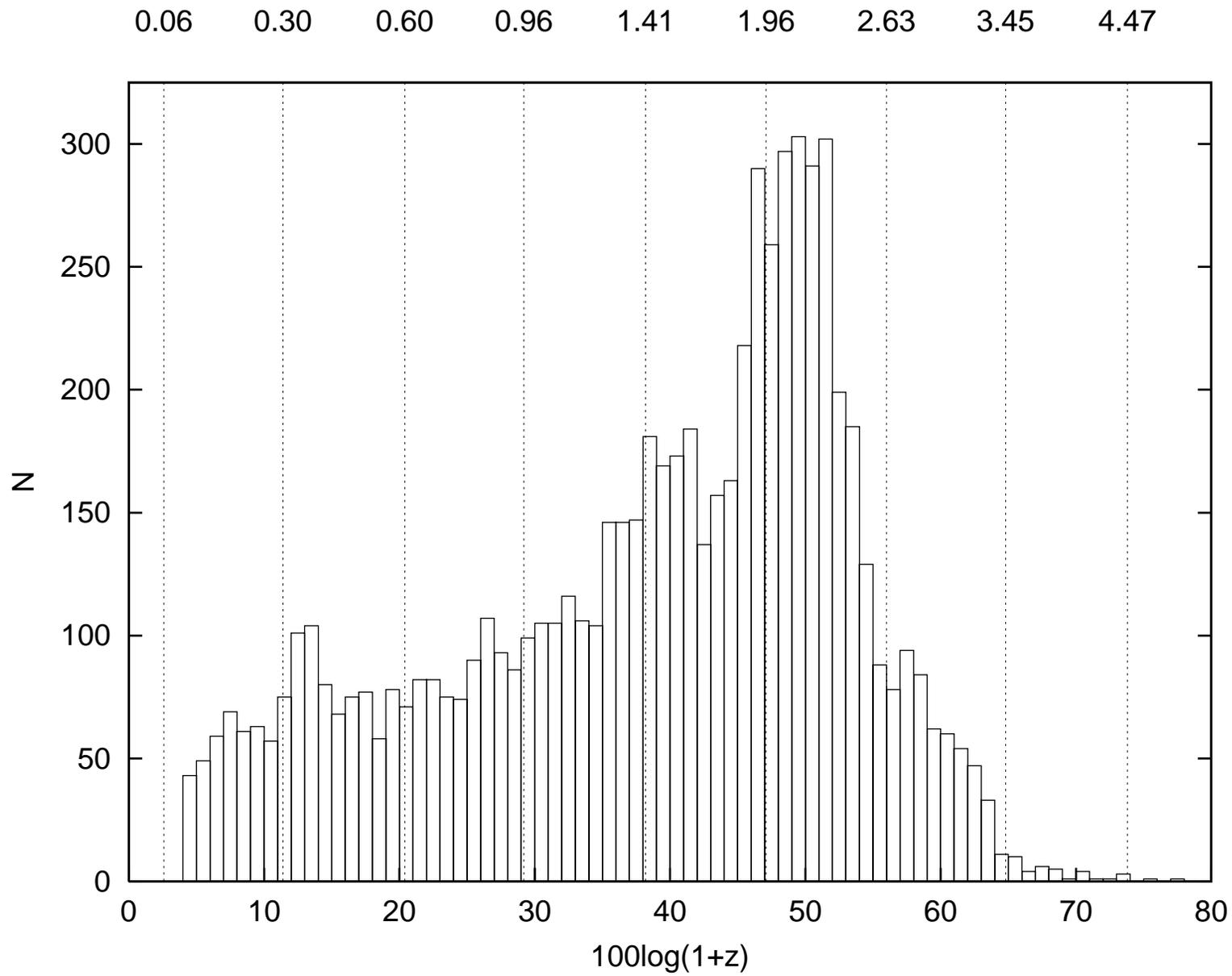
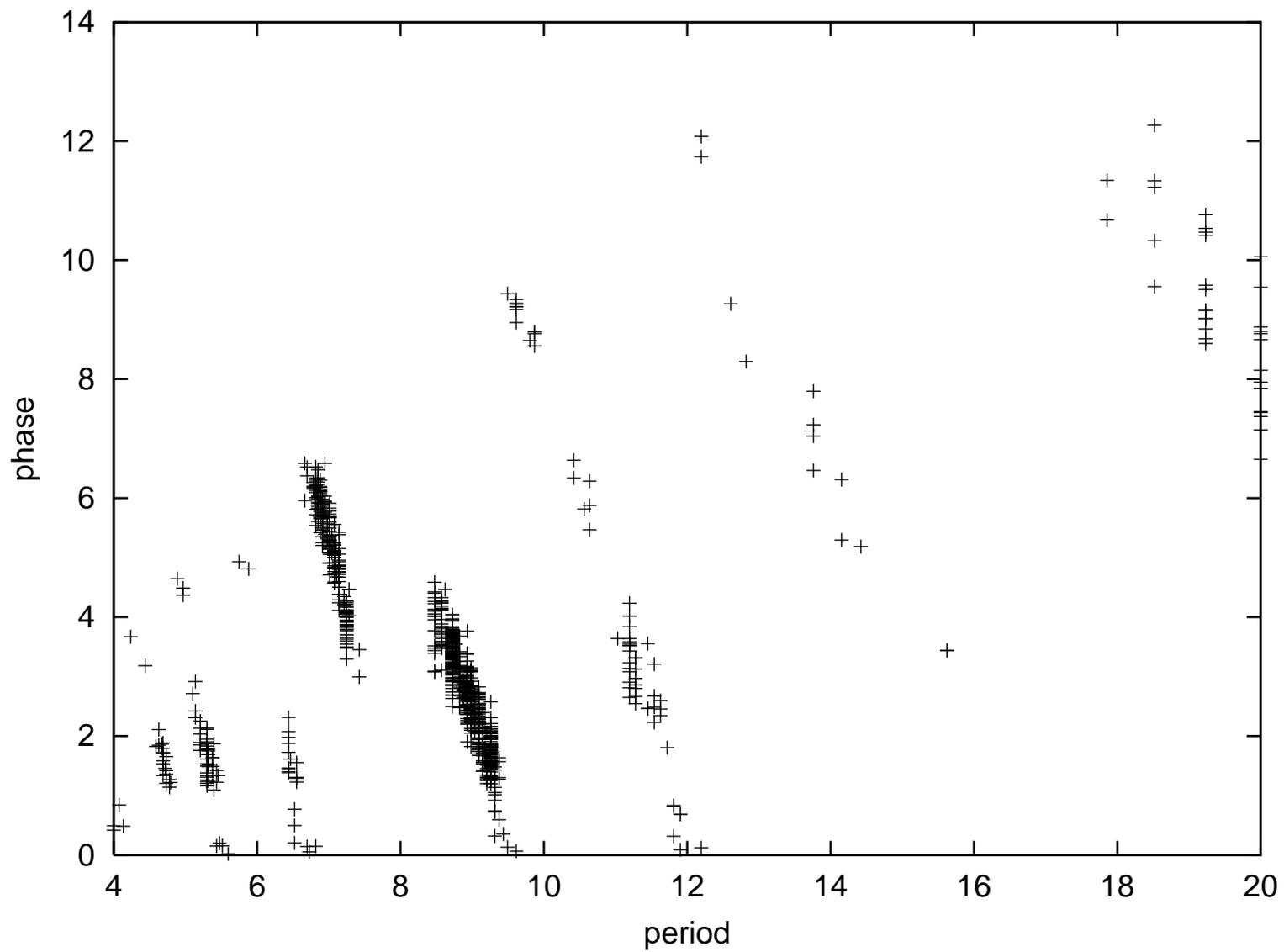
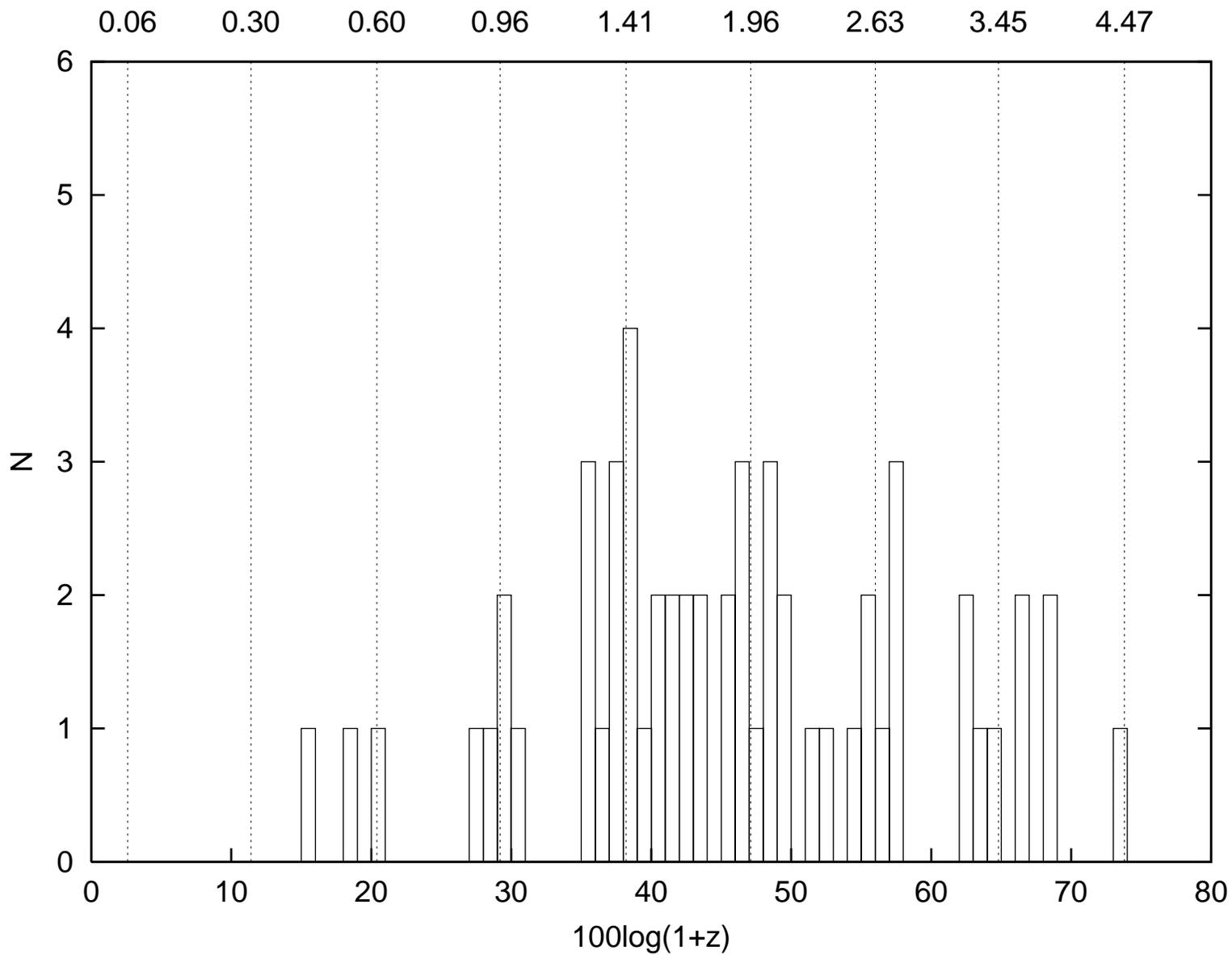
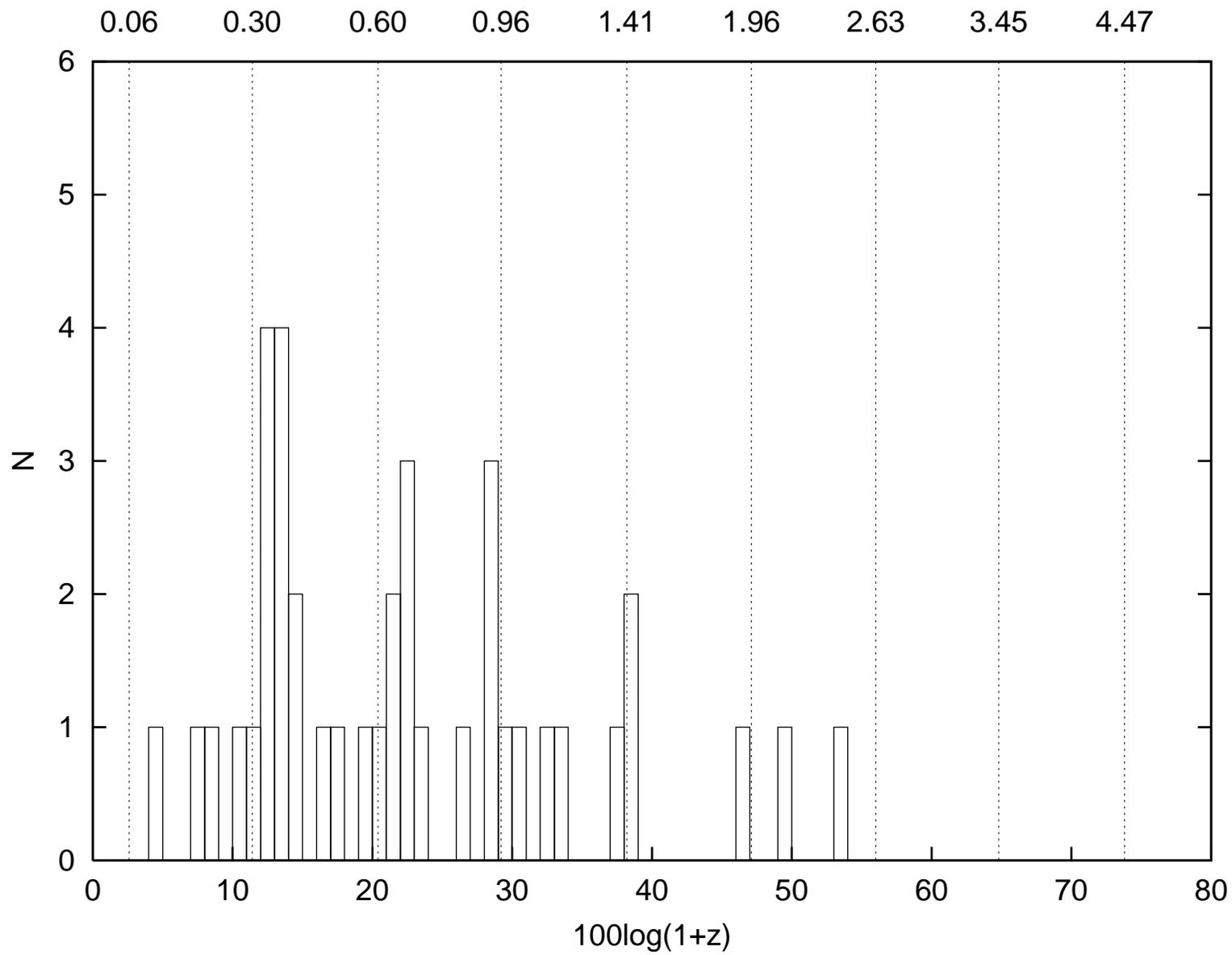
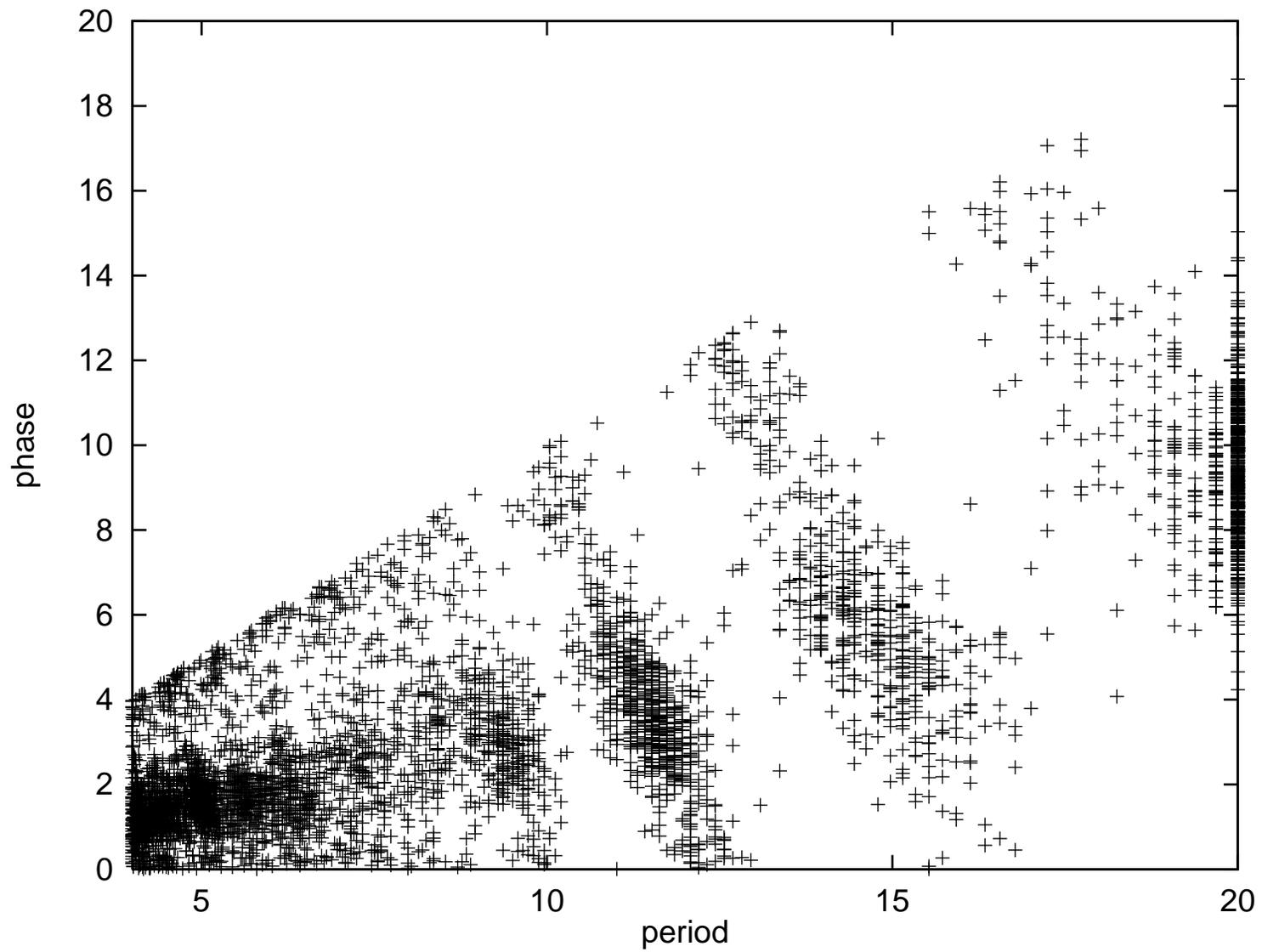
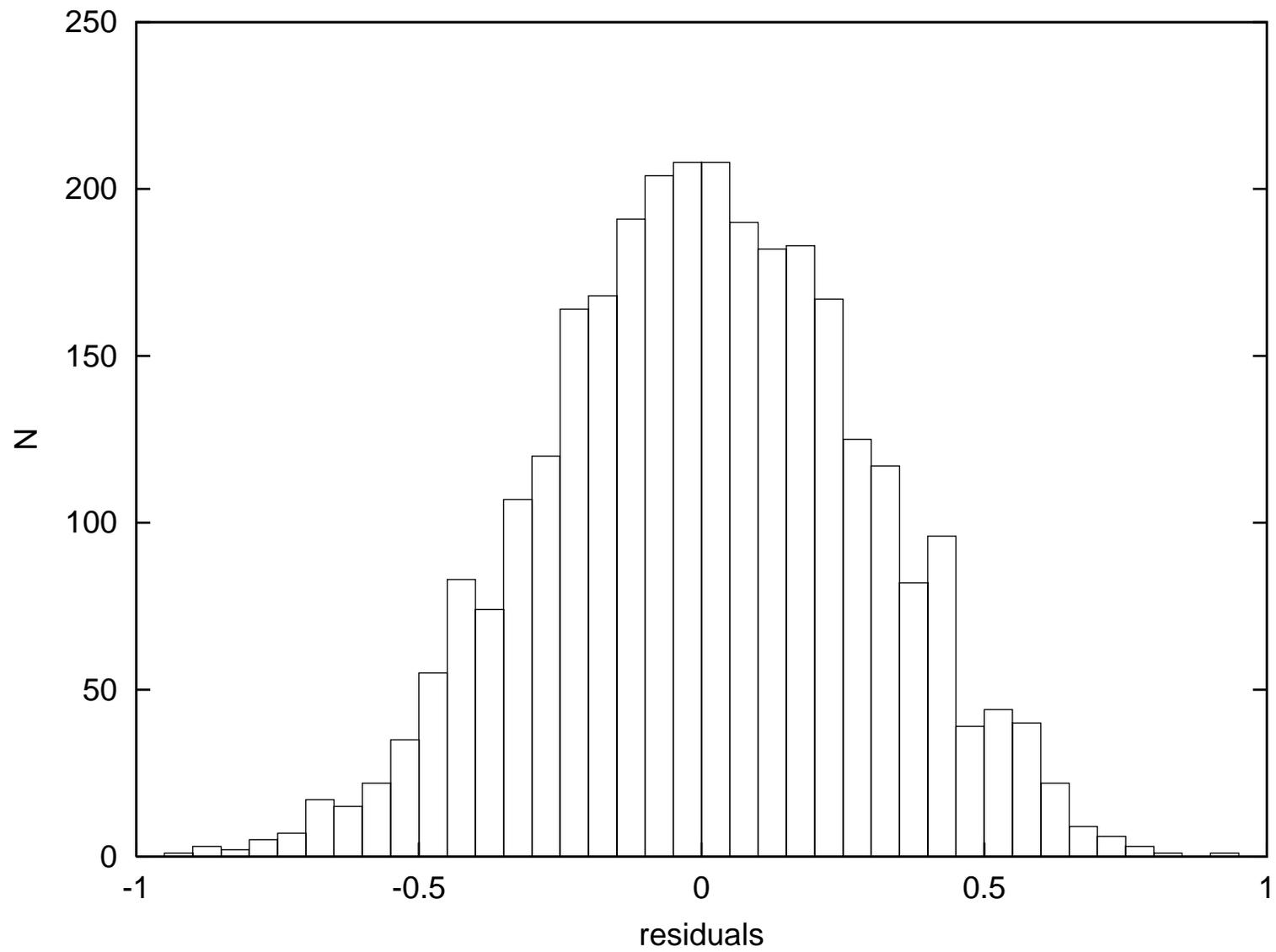
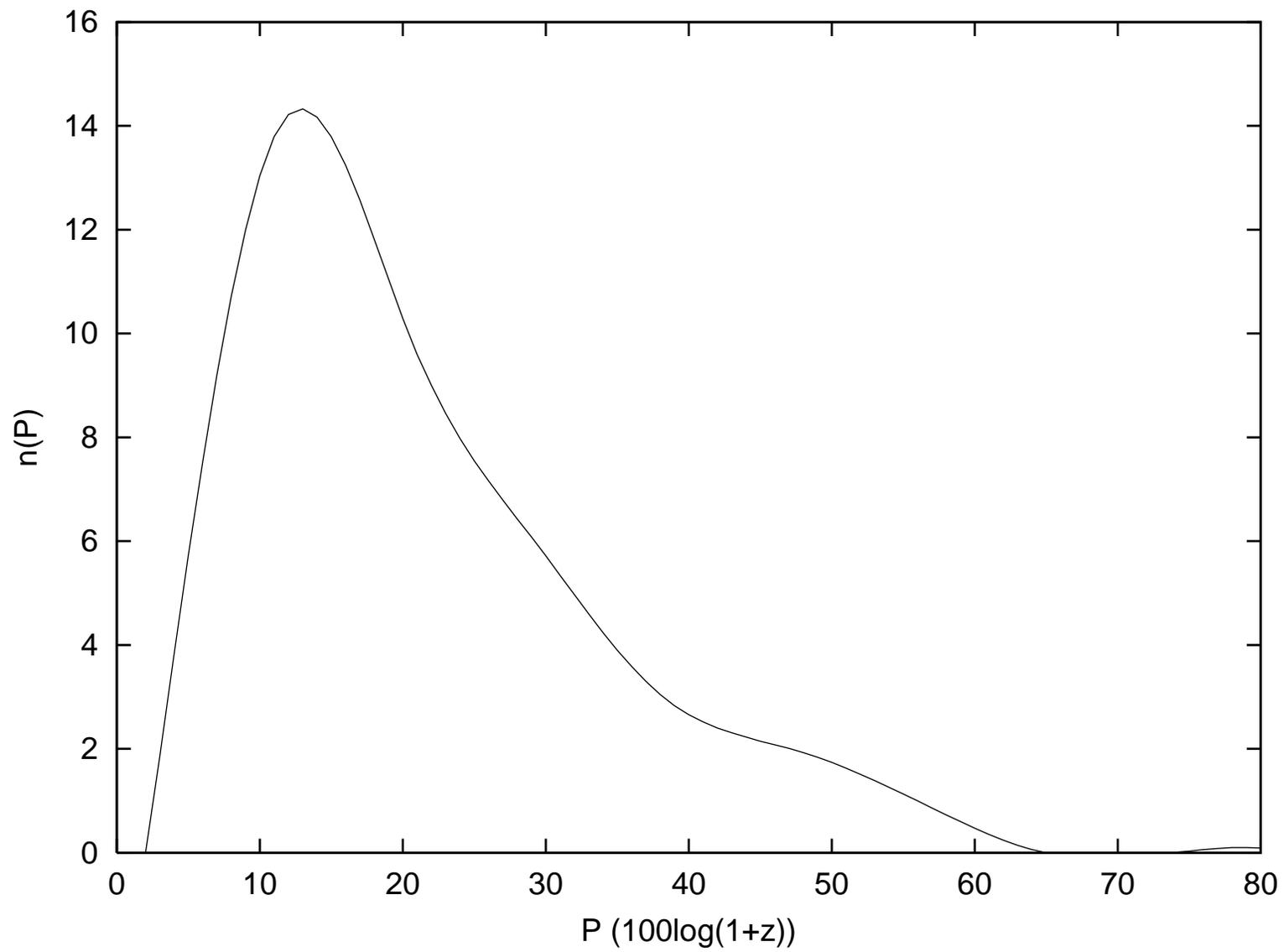
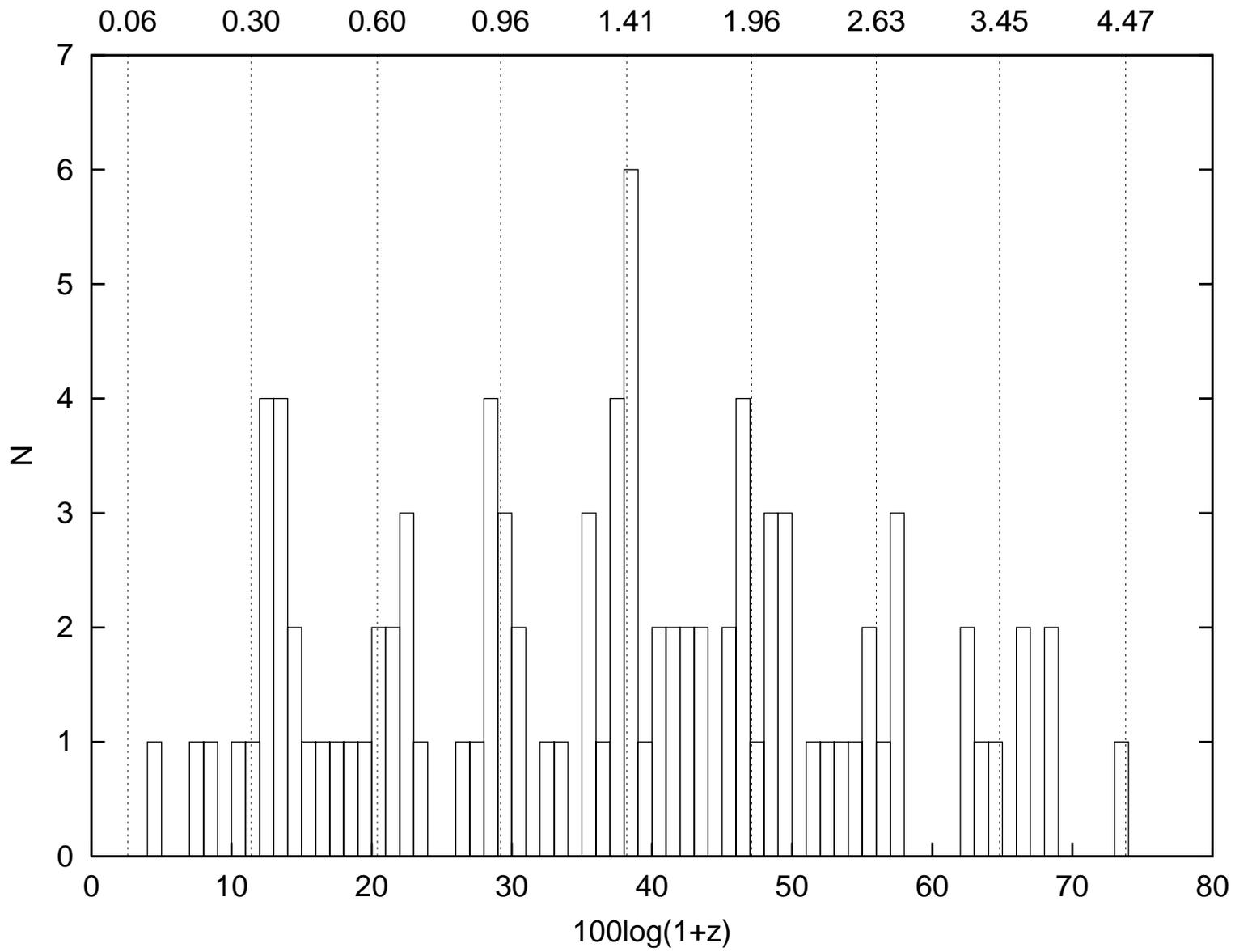
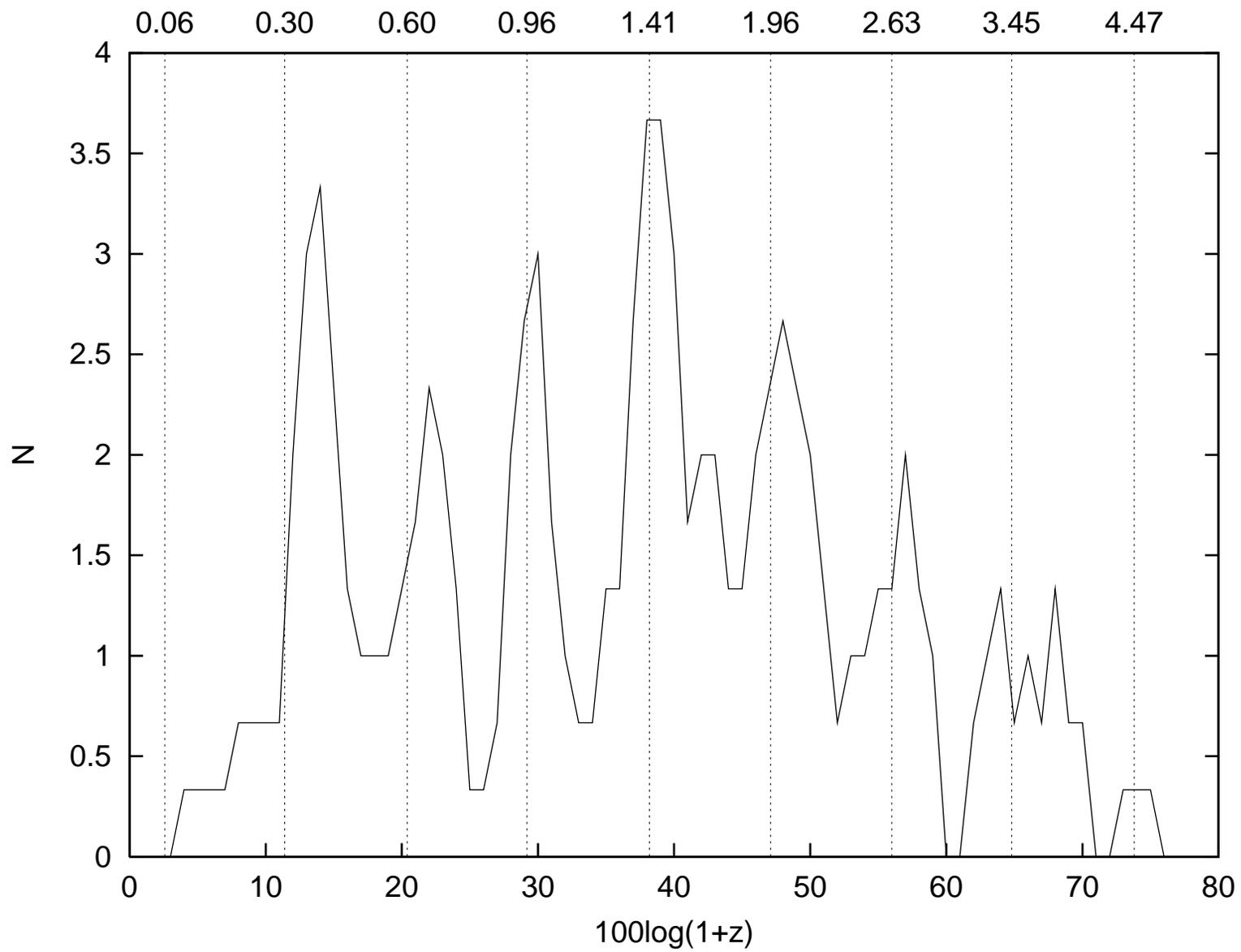
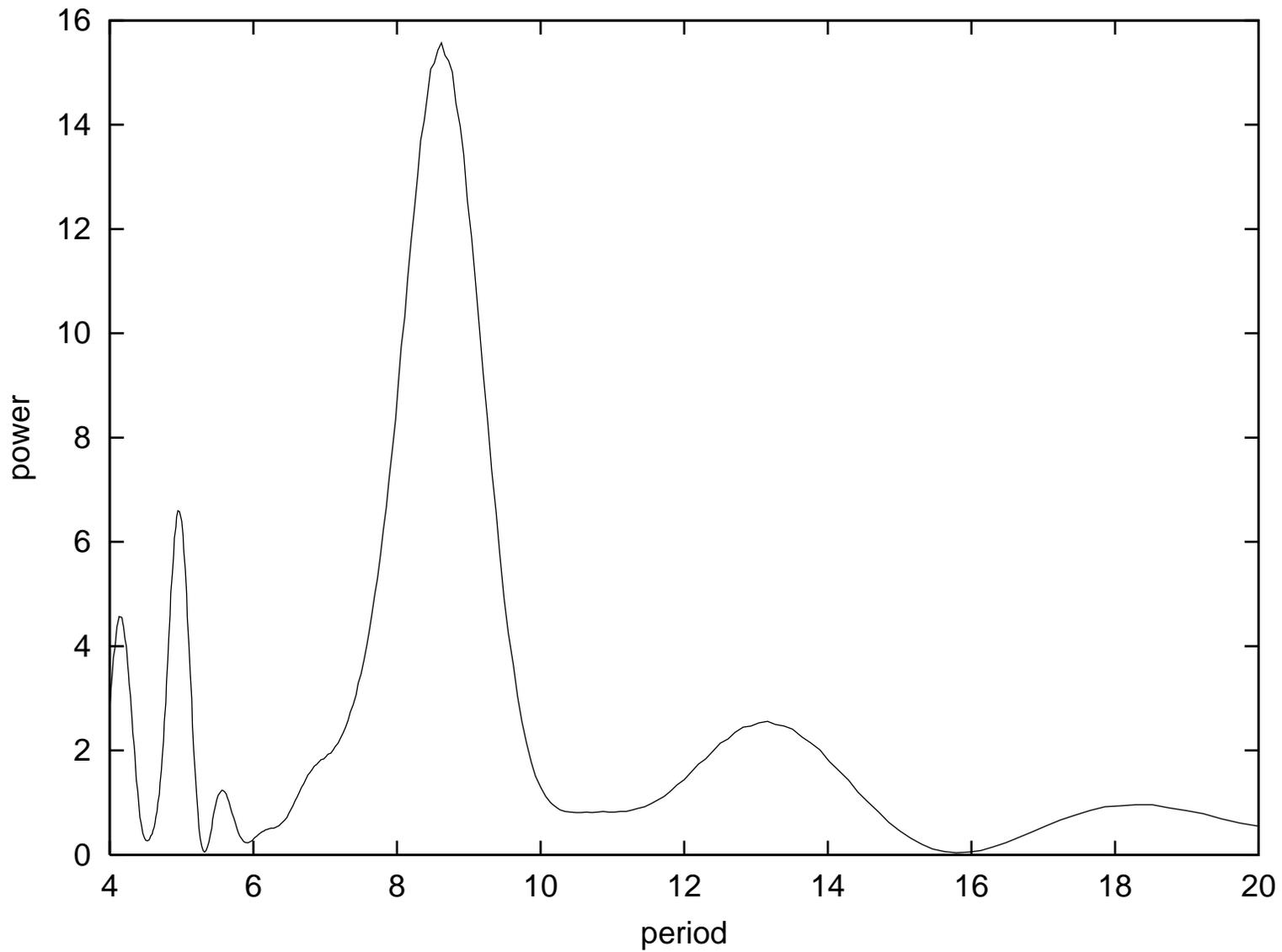
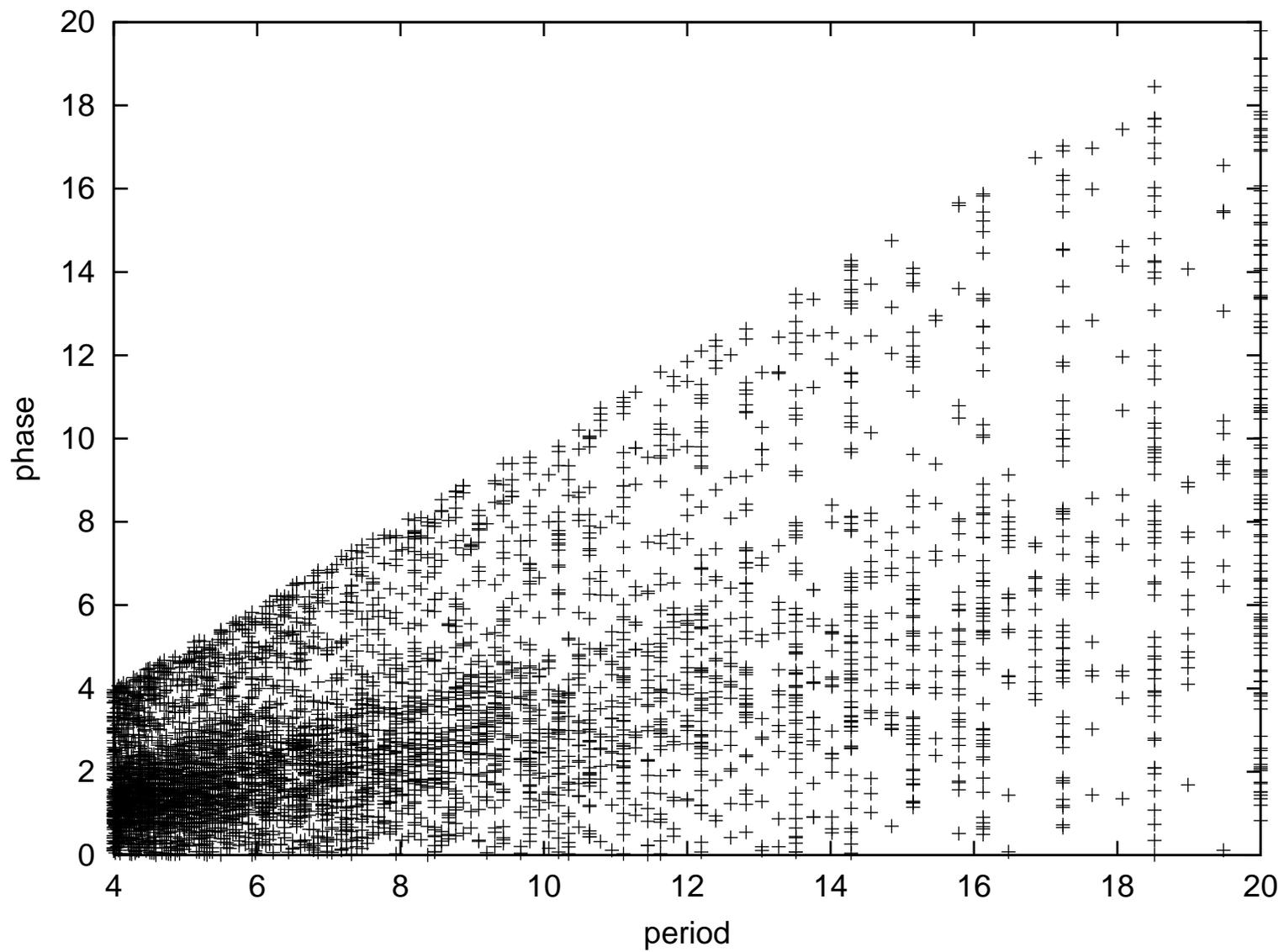
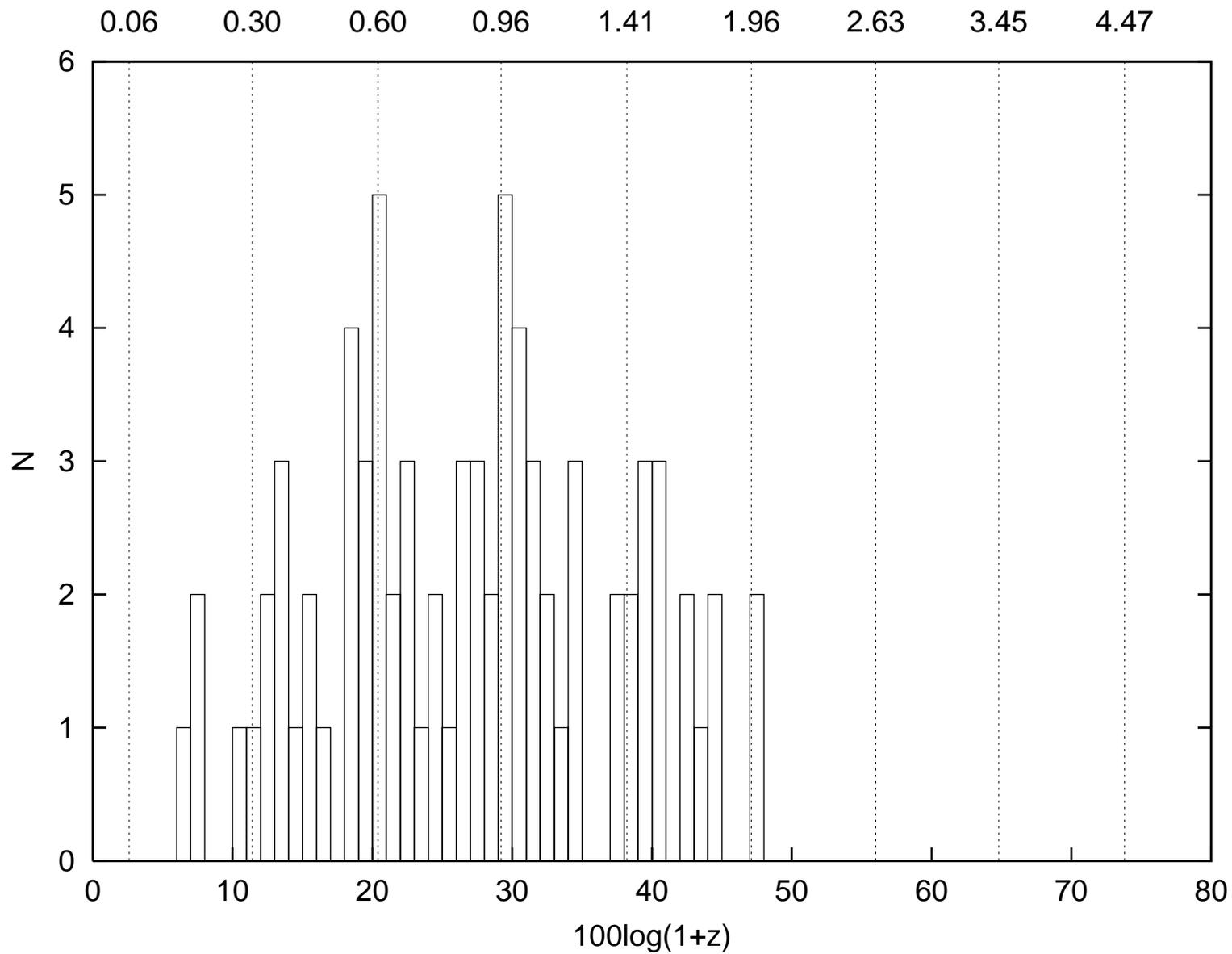
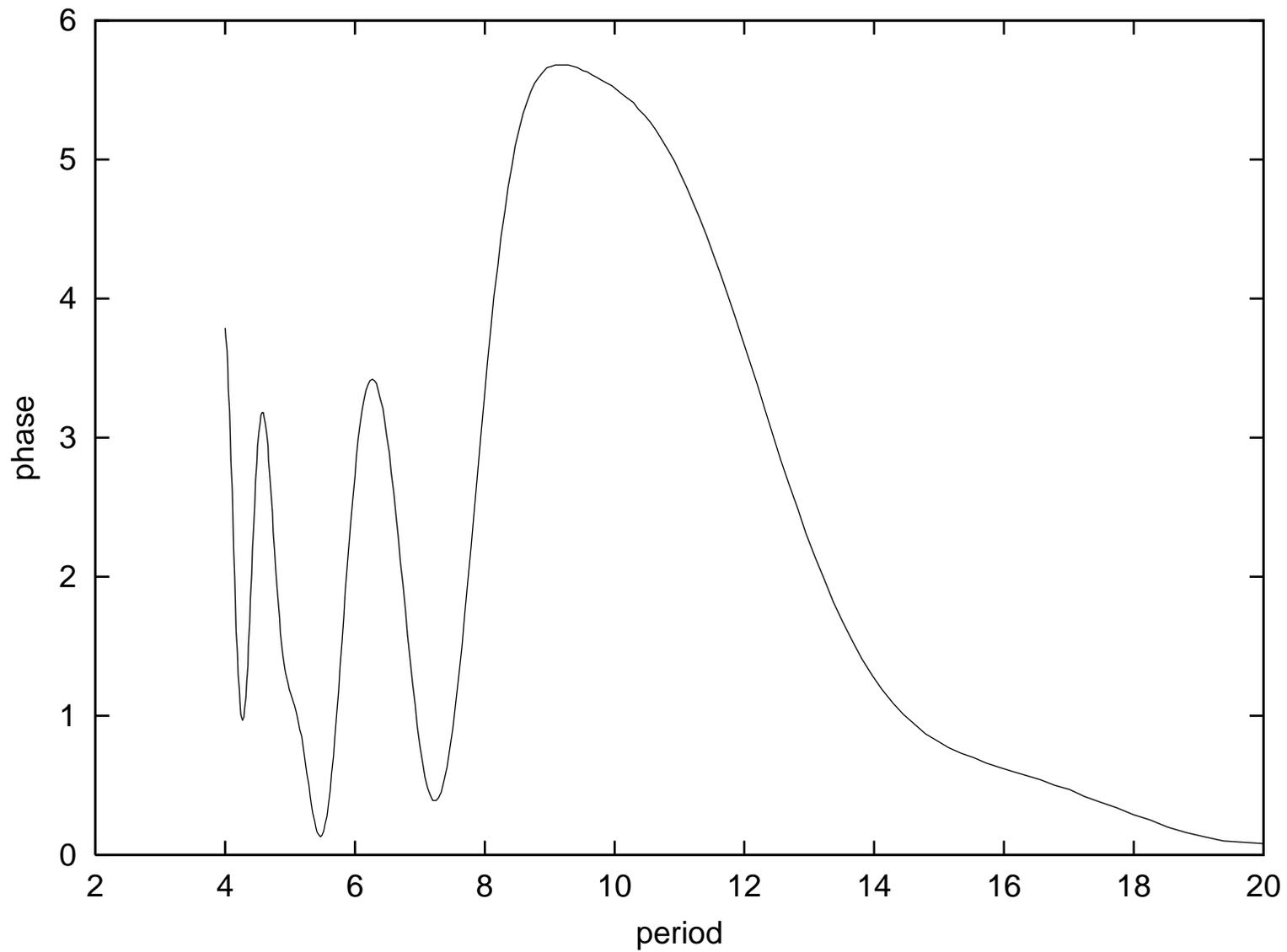
\end{document}